\documentstyle[prcrc,11pt,fleqn,epsf]{article}

\title{New Signatures and Sources for the Detection of WIMP Dark Matter in
the Solar System}

\author{Lawrence M. Krauss\address{Departments of Physics and
Astronomy \\ Case Western Reserve University \\ 10900 Euclid Ave. \\
Cleveland OH 44106-7079}
        \thanks{To appear: Proceedings of Workshop on Sources and Detection
of Dark Matter, UCLA Feb 1998, Research supported in part by DOE,  email:
krauss@theory1.phys.cwru.edu}}

\begin{document}
\maketitle

\begin{abstract}
I first outline new results on the  angular modulation of WIMP
dark matter scattering on targets in terrestrial laboratories, based on 
our uncertainties of the WIMP halo distribution, I then outline an
exciting new result which indicates that for the high end of 
allowed SUSY WIMP scattering cross sections there exists a new
distribution of WIMP dark matter in our solar system which could
produce a dramatically different signal from halo WIMP dark matter in
terrestrial detectors. 
\end{abstract}

\section{Introduction}
If our galactic halo is dominated by WIMPs, searching for signatures which
allow a WIMP signal to be extracted from other backgrounds in 
terrestrial detectors will be of
increasing importance, especially as the predicted signal levels in
detectors decrease.  Here I outline two different ongoing research
programs I am involved in which bear on this issue.  The first involves
the first detailed examination of the directionality of WIMP scattering
in terrestrial detectors for a variety of models of the WIMP distribution
in the galactic halo.  We have developed a formalism which allows the
differential scattering rate over energy and angle to be determined from
any incident distribution, including anisotropic ones.  We then
explore not only the predicted distributions, but also the ability, given
a fixed number of events, to unambiguously demonstrate evidence for
underlying directionality in the events.   The second project
uncovers the existence of a new WIMP distribution in the solar system.
Due to the onset of chaos in solar system orbits induced by Jupiter,
some WIMPs trapped in the solar system due to scattering in the Sun can
move to orbits which no longer intersect the Sun.  We estimate the space
density of such WIMP orbits near the Earth, and find it can be
considerable for realistic SUSY parameters at the upper end of the
allowed range of scattering cross sections. 

\section{Angular Anisotropy and WIMP signatures}

It has been recognized for some time that while the annual modulation
of the WIMP scattering rate in detectors, due to the Earth's motion about
the Sun is small, possible angular anisotropies can be large (i.e.
\cite{sperg}).   However estimates that were performed were based on
the simple assumption of a spherical isothermal sphere model for the
galactic dark matter halo. Morever, while forward-backward assymmetries
can be very large, this measure can be misleading.  Sometimes a large
anisotropy simply reflects a very small number of events, so that
poisson uncertainties can dramatically reduce the significance of any
of such effect.

We have developed a formalism \cite{krausetal} which allows the
incorporation of  aniostropies in the underlying WIMP distribution, due
either  to a non-spherical isothermal distribution, or perhaps due to
non-isothermal flows, for example such as those predicted by
Sikivie \cite{Sikivie}.The rate in a detector depends upon the density
$\rho_{o}$ of WIMPs near the Earth, and the velocity distribution $f(v)$
of WIMPs in the Galactic halo near the Earth.  As a function of the energy
deposited, $Q$, direct-detection experiments measure the number of events
per day per kilogram of detector material.  Qualitatively, this event
rate is simply
$R \sim n\sigma <v> /m_{n}$, where the WIMP number density is
$n={{\rho_{o}} \over {m_{\chi}}}$.
$\sigma$ is the elastic-scattering cross section, 
$<v>$ is the average speed of
the WIMP relative to the target, and we divide the detector mass 
$M_{det}$ by
the target nucleus mass $m_{n}$, to get the number of target nuclei.

More accurately, one should take into account the fact that the WIMPs move in
the halo with velocities determined by $f({\vec v})$,
that the differential cross section depends upon 
$f({\vec v})$ through a form factor ${{d\sigma}\over{d|q|^2}}
\propto F^{2}(Q)$, and that detectors have a threshold energy $E_T$, below
which they are insensitive to WIMP-nuclear recoils.  In addition, the Earth
moves through the Galactic halo and this motion should be taken into account
via $f({\vec v})$.

Consider a WIMP of mass $m_{\chi}$ moving with velocity ${\vec v} = cos\alpha
{\hat x}+sin\alpha sin\beta {\hat y} + sin\alpha cos\beta {\hat z}$ in the
laboratory frame.  This WIMP scatters off  of a nucleus of mass $m_n$ which
recoils with energy $E_{n} = {{m_{n}m_{\chi}^2}\over {(m_{n}+m_{\chi})^2}}
v^{2} (1-\mu)$, where $\mu=cos{{\theta}^{\ast}}$ and ${\theta}^{\ast}$ is the
WIMP scattering angle in the center-of-mass frame.  The velocity $\vec u$ of
the nuclear recoil makes an angle $\gamma$ relative to the $x$ axis:
$cos{\gamma}={{{\hat x}\cdot {\vec u}}\over {|{\vec u}|}}={({{1-\mu}\over 
2})^{1\over 2} cos\alpha - ({{1+\mu}\over 2})^{1\over 2} cos\xi sin\alpha}$,
where $-\sqrt{{1+\mu}\over 2}sin\xi = 
{{{\vec u}\cdot ({\hat x} \times {\vec v})}\over {|{\vec
u}\cdot ({\hat x} \times {\vec v})|}}$. 

The WIMP distribution function $f(v,\alpha,\beta)$ determines the event
 rate
per nucleon in the detector:
$dR = f(v,\alpha,\beta)\,v^{3}dv\,d cos\alpha\,d\beta\,{{d\sigma}\over 
{d\mu}}
d\mu\,{{d\xi}\over {2\pi}}$, 
where $0<v<\infty, -1<cos\alpha<1,0<\beta<2\pi,-1<\mu<1$, 
and $0<\xi<2\pi$.  Assume that the nucleon-WIMP scattering
has following form: ${{d\sigma}\over {d\mu}} = {{\sigma_o}\over 2}F^{2}(Q)$ 
with form factor suppression. The rate of events in which the nucleus
recoils with energy $Q$  is then given by
$
{{d\,R}\over {dQ}}={{\sigma_o \rho_{\chi}}\over {2 m_{r}^{2} m_{\chi}}}
F^{2}(Q){\int_{v_{min}}^{\infty} v\,dv}\>{\int {d\Omega_{\alpha,\beta}}}
\,f(v,\alpha,\beta)
$,
where $v_{min}^{2}={{{(m_{\chi}+m_{n})}^{2}Q}\over {2 m_{\chi}^2 m_{n}}}$ is
the minimum WIMP velocity that can produce a nuclear recoil of energy $Q$.

We then derived the
following equation for the angular differential cross section.
\begin{equation}
{{d\,R}\over {d\Omega_{\gamma,\phi}}}={{\sigma_{o} \rho_{\chi}}
\over {\pi m_{n} m_{\chi}}}
\int_{v_{min}}^{\infty} v^3\,dv F^{2}(Q(v,J))
\,\> {\int d\Omega_{\alpha,\beta}}
\> f(v,\alpha,\beta)\,J(\alpha,\beta;\gamma,\phi)\,\Theta(J)
\end{equation}

Here the function $J(\alpha,\beta;\gamma,\phi)$ takes care of the
arbitrary dependence on the detector position and has the following
property.

\begin{eqnarray}
J(\alpha,\beta;\gamma,\phi)\,\equiv \,[cos\gamma\,cos\alpha\> + 
\> sin\gamma\,sin\alpha\,cos(\phi - \beta)]  \nonumber
\int d\Omega_{\alpha,\beta} J(\alpha,\beta;\gamma,\phi)\,
\Theta(J(\alpha,\beta;\gamma,\phi))\> = \> \pi
\end{eqnarray}

Note that the equation above holds for any $(\gamma,\phi)$ (and
$\Theta(J)$ is the usual step function).

Equation (1) was originally obtained by the series of Jacobians
evaluated for successive transformations of angles from the incident 
direction to the outgoing direction of the recoiled target nucleus.
The end result is actually independent of the intermediate two body
collision channels and gives a simple geometric identity.  When we follow
the track of recoiled particles, not the scattered one as in traditional
fixed target experiments, we have this simple relation between the 
incident angle and the outgoing events.

We have also derived the event rate as a function of both
deposited energy and the outgoing angle.

\begin{equation}
{{d\,R}\over {dQ\,d\Omega_{\gamma,\phi}}}={{{m_{n}} \sigma_{o} \rho_{\chi}}
\over {8 \pi\,m_{r}^4 m_{\chi}}}\,Q\,F^{2}(Q){\int d\Omega_{\alpha,\beta}}
\> f(v(Q,J),\alpha,\beta)
\,{{\Theta(J(\alpha,\beta;\gamma,\phi))}\over {J^{3}}}
\end{equation}

With these three event rate formulae, we have evaluated 
the differential cross sections and event rates for arbitrary
forms of the WIMP halo distribution function. 
\begin{figure}[htb]
\begin{minipage}[t]{80mm}
\epsfxsize = \hsize \epsfbox{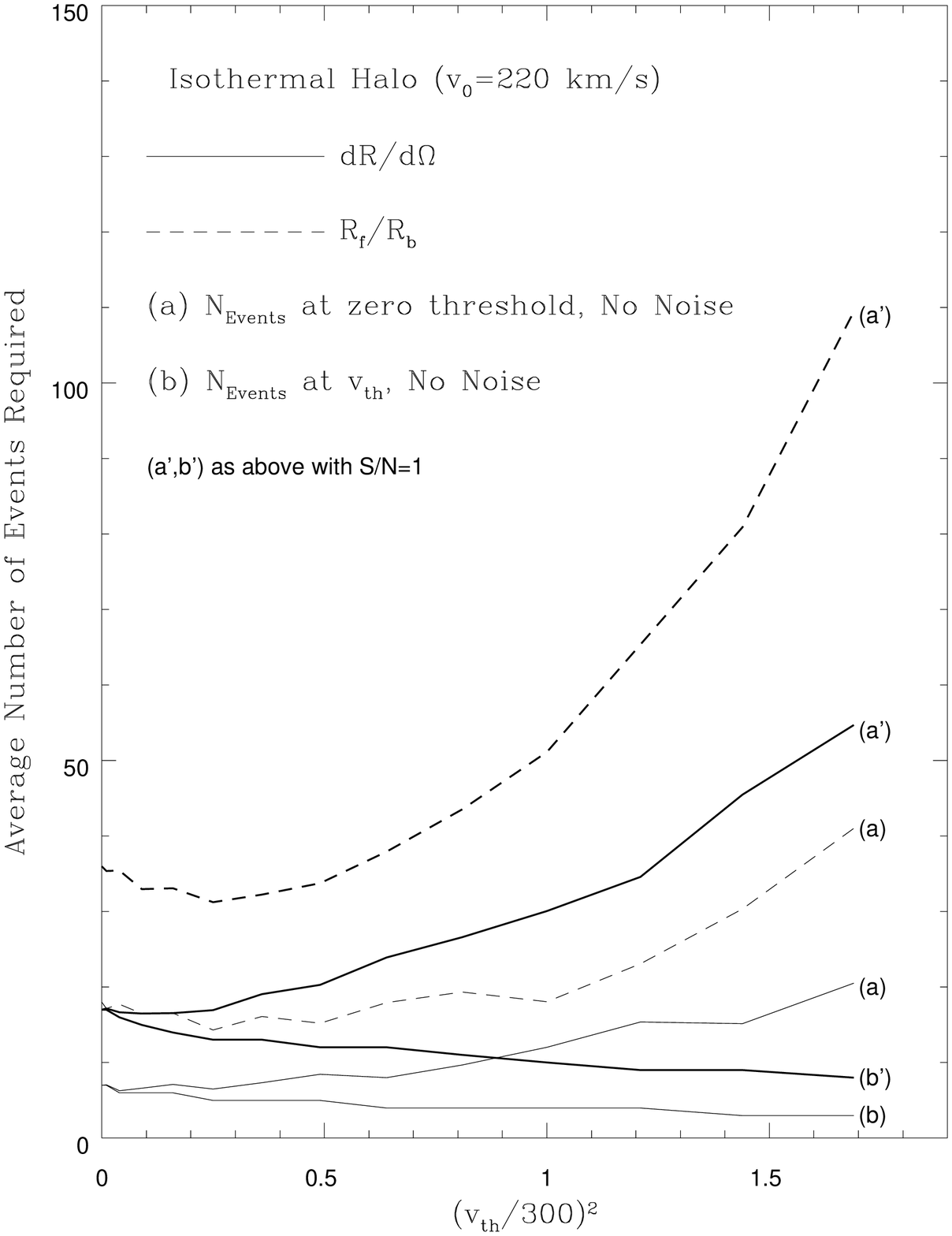}
\caption{Number of Events needed to distinguish halo events arising
from an isothermal
distribution (with $ v_o =220 {\rm kms}^{-1}$) from a flat noise
background}
\end{minipage}
\hspace{\fill}
\begin{minipage}[t]{77mm}
\epsfxsize = \hsize \epsfbox{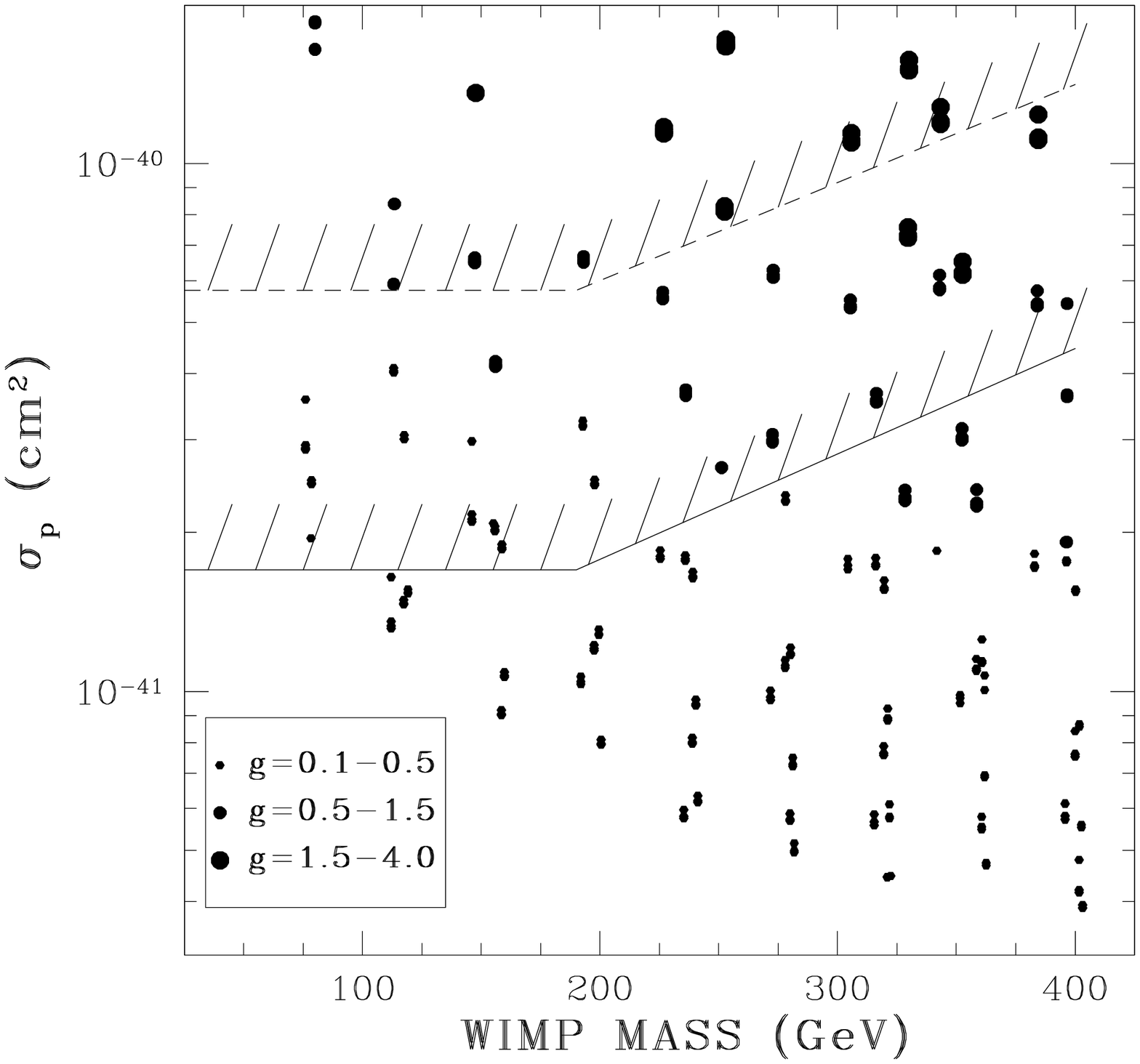}
\caption{Solar capture factor $g_{\rm tot}^{(-10)}$ 
as a function of WIMP mass and 
cross section. Hatched lines give 
existing upper limits assuming a local halo density of
$0.3 {\rm GeV cm^{-3}}$ (lower) or $ 0.1{\rm GeV cm^{-3}}$ (upper). }
\end{minipage}
\end{figure}
  Our results 
\cite{krausetal} indicate that
depending upon the actual distribution function, only 15-30   
events will be required over the course of a year in order to
unambiguously differentiate a non-isotropic distribution from 
a flat background, with a signal to
noise ratio of unity, independent of the
halo distribution, unless it is co-rotating. 
 Displayed
in Figure 1, for example, are the estimates, as a function of
detector threshold, for the number of events required to
differentiate isothermal halo events from a flat background, with and
without noise.  For each set of curves, the lower one gives the number
of events needed above the threshold, while the upper curve gives
the equivalent total number of events required for zero threshold to yield
the required number above the stated threshold.
  Also shown are
the number of events required if only the forward/backward ratio and not
the full distribution is
measured.
With a far greater number of events, features of
the underlying WIMP distribution could be inferred.

\section{A New Solar System WIMP Distribution?}

WIMP dark matter can
scatter elastically in  the Sun, be gravitationally captured,  and
eventually
settle in the  Solar core and annihilate.  Recently however,
we \cite{krausstib1,krausstib2} have found
that perturbations due to the planets, combined with 
the non-Coulomb nature of the
gravitational potential inside the Sun, imply that WIMPS which
are gravitationally captured by scattering
in surface layers of the Sun can evolve chaotically into orbits
whose trajectories no longer intersect the Sun.
  For orbits having a semi-major
axis smaller than 1/2 of Jupiter's orbit, WIMPS can persist in the solar
system for billions of years.  There can thus be a new, previously
unanticipated, distribution of WIMPs which intersect the Earth's orbit.   
For
WIMPs which might be detected in the next generation of underground detectors,
this solar system distribution could be significant, providing a
complementary signal to that of galactic halo dark matter.

We focus on the sub-population of WIMPs which scatter on a nucleus
located near the surface of 
the Sun, and  thereby loose just enough energy to stay in Earth-crossing
orbits.  These will be susceptible to small
gravitational perturbations by the planets.  We are interested then in
the differential capture rate, per energy, and per angular momentum, of
WIMPs in the Sun, and in particular only in the fraction of WIMPs which
have angular momenta in a small range 
$[J_S - \epsilon , J_S]$ 
where $J_S$ is the angular momentum for a WIMP exactly grazing the Sun.
Using the scattering cross section of WIMPs on nuclei given above, 
 one can, after
considerable computation (see \cite{krausstib2}), derive 
the rate with 
which WIMPs scatter on nuclei with atomic number $A$ to end up into bound 
solar orbits with semi-major axis between 
$[a , a + da]$ (corresponding to 
$[\alpha , \alpha + d \alpha]$ with $\alpha \equiv G_N \, M_{Sun} / a$), and
with  specific angular momentum $J \geq J_{\min}$.
\begin{equation}
\left. \frac{d \dot{N}_A}{d \, \alpha} \right\vert_{J \geq J_{\min}} 
\simeq \frac{n_X}{v_o} \int_{r \geq r_{\min}} d^3 {\bf x} \, n_A ({\bf 
x}) \, \sigma_A \left( 1 - \frac{J_{\min}^2}{r^2 \, v_{\rm esc}^2 (r)} 
\right)^{1/2} K_A (r,\alpha) \, . \label{eq2.20}
\end{equation}

Here, the minimum radius
$r_{\min}$  (impact parameter) is defined in terms of the minimum angular
momentum 
$J_{\min}$ by $r_{\min} \, v_{\rm esc} (r_{\min}) \equiv J_{\min}$,
where $v_{\rm esc} (r_{\min})$ is the escape velocity at $r_{\min}$,  
$n_{A,X}$ is the density of atomic targets $A$, and WIMPs $X$,
respectively, $v_o$ is the rms circular velocity of WIMPs in
the Galaxy, and the ``capture function"
$K_A (r,\alpha)$ involves an integral over the WIMP local
phase space distribution at the Sun, weighted over the scattering
form factor.

The Sun scattering events create a 
population of solar-system bound WIMPS, moving (for $ a \sim 1 \, {\rm 
AU}$) on 
very elliptic orbits which traverse the Sun over and over again. For the 
values of WIMP-nuclei cross sections we shall be mostly interested in 
here (corresponding to effective WIMP-proton cross sections in the range 
$4 \times 10^{-42} - 4 \times 10^{-41} \, {\rm cm}^2$), the mean opacity 
of the Sun for orbits with small impact parameters is in the range 
$10^{-4} - 10^{-3}$. This means that after $10^3 - 10^4$ orbits 
these WIMPs will undergo a second 
scattering event in the Sun, and end up in its core where they will
ultimately annihilate. 

The only  way to save some of these WIMPs is to consider a fraction of
WIMPs which have impact parameters
$r_{\rm min}$ in a small  range near the radius of the Sun $R_S$.
Focussing on such a subpopulation of WIMPs has two  advantages: (i) they
traverse a small fraction of the mass of the Sun and therefore their
lifetime on such grazing orbits is greatly increased, and, {\it more
importantly} (ii)  during this time, gravitational perturbations due to
the planets can build up and push them on orbits which no longer cross
the Sun. 

The crucial point to realize is the following. If we consider a WIMP orbit 
with a generic impact parameter $r_{\rm min} << R_S$, it will 
undergo a large perihelion precession $\Delta \, \omega \sim 2\pi$ per orbit, 
i.e. $\dot \omega \sim n$, because the potential $U(r)$ {\it within the 
Sun} is modified compared to the exterior $1/r$ potential leading  to the
absence of perihelion motion. In other words, the  trajectory of the WIMP will
generically be a fast advancing {\it rosette}.  When the rosette
motion is fast, planetary perturbations do not induce any secular evolution in
the semi-major axis and in the  eccentricity of the WIMP orbit. Such
WIMPs will end up in the  core of the Sun.

A new situation arises for WIMP orbits which graze the Sun, because these 
orbits feel essentially a $1/r$ potential due to the Sun, so that their 
rosette motion will be very slow.  One can then show that planetary
perturbations will in this case alter the eccentricity of orbits so
that such orbits will no longer intersect the Sun.

We can finally estimate the density of WIMPs which diffuse out to solar-bound
orbits and which can survive to the present time by simply integrating our
differential capture rate over all trajectories $J > J_{min} \equiv G_{min}$
which end up out of the Sun, suitably averaged over the initial WIMP
distribution.  The rate (per $\alpha = G_N \, m_S / a$) of solar capture of
WIMPs which  subsequently survive out of the Sun to stay within the inner solar
system then depends on the $A$-dependent combination:
$ g_A \equiv \frac{f_A}{m_A} \, \sigma_A \, K_A^s $, where $f_A$ is the
fraction (by mass) of element $A$ in the Sun, and $ K_A^s $ is the
Sun-surface value of the capture function mentioned above.
Note that the $A$-dependence is entirely contained in  $g_A$ 
with dimensions $[\hbox{cross section}] / [\hbox{mass}]$.  The total capture
rate is then dependent on ${\displaystyle \sum_A} 
\, g_A  =g_{\rm tot}$.
From the point of view of experiment, however, 
the differential rate, $dR / dQ$, per keV per kg per day, of scattering events
in a laboratory sample made of element $A$ is what counts.  Comparing this
differential rate due to our new WIMP population with that due to the
predicted galactic halo population we find a ratio which reaches the 
maximum,

\begin{equation}
\rho (Q) \equiv \frac{(dR / dQ)^{\rm new}}{(dR / dQ)^{\rm standard}}
\  = 1.1 \, g_{\rm tot}^{(-10)} \, ,
\end{equation}
for energy transfers $Q$ smaller than
$Q_E = 2 (m_X/(m_X + m_A))^2 m_A v_E^2$, where
$g_{\rm tot}^{(-10)}$ is the value of $g_{\rm tot}$ when cross sections
are expressed in terms of $10^{-10} GeV^{-2}$. For a target of Germanium
(the present material of choice in several ongoing cryogenic
WIMP detection experiments), $Q_E$ is in the keV range.
If $g_{\rm tot}^{(-10)} \sim 1$  this yields a 100~\% increase of 
the differential event rate below the value $Q = Q_E \sim {\rm keV}$ which is
typical for the energy deposit due to the new WIMP population, whose
characteristic velocity $\sim v_E$ is smaller than that
of galactic halo WIMPs.

This is our central result. We expect that this is in fact an
underestimate of the actual contribution due to WIMPs trapped in the solar
system, because we have incorporated in our calculations only the 
average long-range perturbing effects of the planets. 
 In any case, the relevance of this result
depends completely on the actual value of $g_{\rm 
tot}^{(-10)}$.  Interestingly, if one explores the allowed parameter space of
neutralinos in the Minimal Supersymmetric Standard Model, shown for example
in Figure 2, for SUSY parameter $\mu >0$, we find 
values of $g_{\rm tot}^{(-10)}$ in excess of 1 for
the range of parameters which involve a remnant WIMP density in excess of about
$2 \%$ of the closure density, and whose scattering rate remains below the
sensitivity of current detectors.

Could such a new WIMP population be detectable?  Existing detectors tend to
lose sensitivity in the range of a few keV.  However, these results
motivate pushing hard in this direction. In the first place,
the new population will have a strongly anisotropic velocity
distribution.  Not only will this greatly help distinguish it from
backgrounds, but a comparison of the annual modulation of any signal from
this distribution with the higher energy signal 
from a halo WIMP distribution would be striking.  
If neutralinos exist in the range detectable at
the next generation of detectors, this new WIMP population must exist at a
sizeable level.  Finally, we note that the indirect neutrino signature of such
WIMPs which might subsequently be captured by the Earth, and annihilate, could
be dramatic.  The new population has a characteristic velocity which more
closely matches the escape velocity from the Earth than does the background
halo population.  As a result, resonant capture off elements such as Iron in
the Earth could be greatly enhanced.

\end{document}